\documentclass[twocolumn]{revtex4}
\usepackage{hyperref}
\usepackage{amsmath,amsfonts}
\begin{document}

\title{{Finite BRST Mapping in Higher Derivative Models }}
 \author{Pavel Yu. Moshin${}^{a}$}
  \email{moshin@phys.tsu.ru} 
   \author{Sudhaker Upadhyay${}^{b}$}
  \email{sudhakerupadhyay@gmail.com}  
   \author{Ricardo A. Castro${}^{c}$}
 \email{rcastro@if.usp.br}
    
 \affiliation{${}^{a}$
              Faculty of Physics, National Research Tomsk State University, 634050,
Tomsk, Russia}
  \affiliation{${}^{b}$  Centre for Theoretical Studies, Indian Institute of Technology
Kharagpur, Kharagpur-721302, WB, India}
 \affiliation{${}^{c}$Institute of Physics, University of S\~{a}o Paulo, Rua do Mat\~{a}o
1371, CEP 05508-090, Cidade Universit\'{a}ria, S\~{a}o Paulo, Brazil}

\begin{abstract}
We continue the study of finite field dependent  BRST (FFBRST) symmetry in
the quantum theory of gauge fields. An expression for the Jacobian of path
integral measure is presented, depending on a finite field-dependent
parameter, and the FFBRST symmetry is then applied to a number of
well-established quantum gauge theories in a form which includes
higher-derivative terms. Specifically, we examine the corresponding versions
of the Maxwell theory, non-Abelian vector field theory, and gravitation
theory. We present a systematic mapping between different forms of
gauge-fixing, including those with higher-derivative terms, for which these
theories have better renormalization properties. In doing so, we also
provide the independence of the S-matrix from a particular gauge-fixing with
higher derivatives. Following this method, a higher-derivative quantum
action can be constructed for any gauge theory in the FFBRST framework.
\end{abstract}
\maketitle

\section{Introduction}

Higher-derivative (HD) field theories naturally emerge, due to various
reasons, as effective theories in a wide area of physics. Perhaps the best
known example is gravity, in which higher-order terms in the curvature arise
either from underlying string dynamics, or from quantizing matter fields.
Quite often, HD terms are added to a given standard theory as corrections.
In gravity theories, HD terms ensure renormalizability \cite{stelle}.
Besides the renormalization properties, the known facts about the theory
  include the particle contents, given by the linear decomposition
of the HD propagator into the parts containing second-order poles. 
 Some
issues related to the equations of motion have also been discussed \cite{gib}. 
Unitarity in renormalizable HD quantum gravity has been examined, and the
presence of a massive spin-2 ghost in the bare propagator is found to be
inconclusive \cite{anto}.

On the other hand, in the case of a massive relativistic particle, the
action is extended by the curvature term, being higher-derivative by its
nature. This particle model, introduced quite a long time ago \cite{pisarski}, is still under active consideration \cite{nesterenko,plyuschay1,plyuschayA,plyuschay2,ramos,ramos2,BMP,rbs}. The introduction of HD
fields is not limited to this particular area. Instead, it has been
considered in diverse theoretical models, such as electrodynamics \cite%
{podolsky1,podolsky2}, supersymmetry \cite{Iliopoulos,Gama},
noncommutative theory \cite{clz,plyuschay6}, cosmology \cite{neupane,nojiri4}, extended Maxwell--Chern--Simon theory \cite{reyes,MP}, theory of
anyons \cite{plyuschay3,plyuschay4,plyuschay5}, relativistic particles
with torsion \cite{plyuschay7}, membrane description for the electron \cite%
{cordero,paul}, etc. There are many more gravity models in which HD
corrections are added to the Einstein--Hilbert action \cite{accioly,soti,gullu,ohta}. HD terms acquire relevance also in the context of string
theory \cite{polya,elie}. Thus, the importance of HD terms cannot be
overestimated.

In quantizing gauge field theories, the Becchi-Rouet-Stora-Tyutin (BRST)
formalism \cite{brst,tyu,wei,sudu} provides a comparatively rigorous
mathematical scheme. Even though the BRST formulation is a powerful approach
to quantize gauge theories, which simplifies the study of renormalizability
and unitarity of gauge theories, the implementation of this approach in HD
theories is quite nontrivial and poses problems. Despite this fact, in usual 
gauge field theories the standard BRST symmetry has been generalized by
allowing the transformation parameter to be finite and field dependent \cite%
{sdj}.

Thus generalized BRST symmetry transformations, or so-called  FFBRST  symmetry, lead to a non-trivial Jacobian of
functional measure and find applications in a wide area of gauge theories,
including gravity \cite{sdj,sdj1,susk}. For instance, the celebrated Gribov
problem \cite{gri,zwan,zwan1} has been addressed in the framework of FFBRST
formulation (see Ref. \cite{sb} and references therein). In this article, we
present an elegant approach to derive the Jacobian of functional measure, as
compared to the original study of \cite{sdj}. The advantage of the present
approach is that one has no need to provide an ansatz for a local functional
subjected to some boundary conditions. On top of that, one has no need to
solve differential equations satisfying certain initial boundary conditions
to obtain a precise expression for the Jacobian. Here, the evaluation of a
Jacobian only requires that one provide a suitable infinitesimal
field-dependent parameter.

FFBRST transformations have been given an emphasis in higher-form gauge
theories \cite{smm}. Further, in supersymmetric M-theories \cite{bl,g,abjm}
such developments have also been studied \cite{fs,sudd,fsm}. Recently, the
gravity models have been explored in the context of FFBRST transformations 
\cite{bisu}. Such generalizations are established at the quantum level \cite%
{ssb,sud001}, using the BV technique \cite{ht}. Recently, the FFBRST
formulation has acquired relevance in topological gauge theories \cite{rs}.
Moshin and Reshetnyak, for the first time \cite{ale}, systematically
incorporated BRST-antiBRST symmetry into Yang--Mills theories within the
context of finite transformations that deals with the case of a quadratic
dependence on the transformation parameters. Further, the concept of finite
BRST-antiBRST symmetry in general gauge theories has been used in Refs. \cite%
{ale1,mos1}, whereas Ref. \cite{mos} by the same authors generalizes the
corresponding parameters to the case of arbitrary  Grassmann odd 
field-dependent parameters, as compared to the so-called \textquotedblleft
potential\textquotedblright\ form of parameters \cite{ale,ale1,mos1}.
The generalization of supersymmetry   transformations with $m$ generators  and 
  physical consequences of Grassmann odd transformations  are also studied in Ref. \cite{asm}.

A natural question arises concerning the application of the FFBRST formalism
to HD theories. Indeed, it is not surprising, despite a considerable amount
of research on HD models, that this issue so far remains unstudied. The
basic motivation for this paper is to express FFBRST transformations in a
more transparent way and to explore the possible applications of this
formalism to HD gauge theories. In this context, we make a simplified way to
FFBRST transformations by following Ref. \cite{sdj} up to some good extent.
As originally, we make all the fields parameter-dependent by a continuous
interpolation such that, at one limit, it corresponds to the original field
and, at another limit, to a transformed field. Further, we define an
infinitesimal field-dependent transformation by making the constant
parameter infinitesimally field-dependent. Now, we integrate such an
infinitesimal field-dependent transformation to obtain an FFBRST
transformation. Then, we evaluate the Jacobian of functional measure under
FFBRST with an arbitrary field-dependent parameter. Further, we apply the
resulting FFBRST transformation to various HD models, which leads to some
interesting observations. First, we examine the FFBRST transformation in
Maxwell theory and find that for a particular choice of the field-dependent
parameter it maps gauge-fixing to an HD version of this theory, which also
preserves the independence of the S-matrix from any particular gauge-fixing.
We further apply FFBRST transformations to non-Abelian and gravitational
theories, so as to extend the results and validity of our treatment. Indeed,
we find that this treatment works in each of the gauge theories involved.
Since HD terms play an important part in achieving the renormalization of
ultraviolet (UV) divergent gauge theories, the present technique could be of
help in dealing with UV-divergent gauge theories.

The paper is organized as follows. In Section II, we present the
construction of FFBRST transformations in a simplified way. We derive a
manifest expression for the Jacobian with no need of boundary conditions.
Further, in Section III, we illustrate various HD models and discuss their
BRST quantization. To be specific, in Subsection IIIA, we discuss BRST and
FFBRST transformations in Maxwell theory and its HD version. In this
description, we derive a Jacobian which consists only of BRST-exact terms
for the HD model. In Subsection IIIB, we use FFBRST transformations to
produce an HD non-Abelian action. In Subsection IIIC, we study BRST and
FFBRST transformations in HD gravity. We map HD gravity to its quantum
version through FFBRST transformations. In Section IV, we summarize the
results and suggest some future motivations.

\section{Construction of finite field-dependent BRST transformations}

In this section, we illustrate the FFBRST formulation, on general grounds,
within a simplified approach following Ref. \cite{sdj} up to some good
extent. Let us begin by defining infinitesimal BRST transformations for a
generic field $\phi (x)$ as follows: 
\begin{equation}
\phi (x)\longrightarrow \phi ^{\prime }(x)=\phi (x)+s_{b}\phi (x)\ \Lambda ,
\label{a}
\end{equation}%
where $s_{b}\phi $ is the so-called Slavnov variation, and $\Lambda $ is an
infinitesimal anticommuting parameter with no spacetime dependence. Under
such transformations, the path integral measure remains invariant \cite{wei}.

Now, the field $\phi (x)$ turns into a continuous parameter ($\kappa ;0\leq
\kappa \leq 1$) such that $\phi (x,\kappa =0)=\phi (x)$ is the original
field, and $\phi (x,\kappa =1)=\phi ^{\prime }(x)=\phi (x)+s_{b}\phi
(x)\Theta \lbrack \phi ]$ is an FFBRST-transformed field characterized by a
finite field-dependent parameter $\Theta \lbrack \phi ]$. To justify FFBRST
transformations, we construct the following infinitesimal field-dependent
BRST transformations \cite{sdj}: 
\begin{equation}
\frac{d\phi (x,\kappa )}{d\kappa }=s_{b}\phi (x,\kappa )\Theta ^{\prime
}[\phi (\kappa )],  \label{b}
\end{equation}%
where $\Theta ^{\prime }[\phi (\kappa )]$ is an infinitesimal
field-dependent parameter.

Further, we proceed by making integration over $\kappa $ and arrive at the
following field-dependent transformation \cite{sdj}: 
\begin{equation}
\phi (x,\kappa )=\phi (x,0)+s_{b}\phi (x,0)\Theta \lbrack \phi (\kappa )].
\end{equation}%
Here, $\Theta \lbrack \phi (\kappa )]$ is related to $\Theta ^{\prime }[\phi
(\kappa )]$ through 
\begin{eqnarray}
{\Theta \lbrack \phi (k)]}&=&\int_{0}^{\kappa }d\kappa \ {\Theta }^{\prime
}[\phi (\kappa )],\nonumber\\
&=&{\Theta }^{\prime }[\phi (0)]\frac{\exp {(\kappa f[\phi
(0)])}-1}{f[\phi (0)},
\end{eqnarray}%
with $f[\phi (k)]=\frac{\delta {\Theta }^{\prime }}{\delta \phi }s_{b}\phi $. For the boundary value of $\kappa $ (i.e., $\kappa =1$), this yields the
FFBRST transformation 
\begin{equation}
\delta _{b}\phi (x)=\phi ^{\prime }(x)-\phi (x)=s_{b}\phi (x)\Theta \lbrack
\phi (1)].  \label{c}
\end{equation}%
It is easy to verify that the resulting FFBRST transformations with a
field-dependent parameter also provide a symmetry of the quantum action, but
the price to pay is that these are no longer nilpotent and do not leave the
functional measure invariant. Incidentally, the path integral measure also
changes non-trivially under these transformations, leading to a non-trivial
Jacobian within functional integration. So it is worthwhile to compute an
explicit Jacobian of functional measure under such transformations and
follow the pertaining consequences.

\subsection{Jacobian for finite field-dependent BRST transformations}

In this subsection, we compute the Jacobian for path integral measure under
FFBRST transformations with arbitrary and specific parameters. Let us start
by defining the vacuum functional in Maxwell theory, described by a quantum
action $S_{FP}[\phi ]$, 
\begin{equation}
Z[0]=\int \mathcal{D}\phi \ e^{iS_{FP}[\phi ]},  \label{zen}
\end{equation}%
where $\mathcal{D}\phi $ stands for the complete functional measure.
Furthermore, in order to compute the Jacobian of functional measure under
FFBRST transformations, we observe \cite{sdj} 
\begin{equation}
\mathcal{D}\phi (\kappa )=J(\kappa )\mathcal{D}\phi (\kappa )=J(\kappa
+d\kappa )\mathcal{D}\phi (\kappa +d\kappa ).
\end{equation}%
Because of its infinitesimal nature, the transformation from $\phi (\kappa )$
to $\phi (\kappa +d\kappa )$ can be presented as \cite{sdj} 
\begin{equation}
\frac{J(\kappa )}{J(\kappa +d\kappa )}=\sum_{\phi }\pm \frac{{\delta }\phi
(\kappa +d\kappa )}{{\delta }\phi (\kappa )},
\end{equation}%
where the $+$ sign is used for bosonic fields, and $-$ is used for fermionic
fields. Now, upon making the Taylor expansion, we obtain \cite{sdj} 
\begin{equation}
 \frac{1}{J}\frac{dJ}{d\kappa }  =- \int d^{4}x\sum_{\phi
}\pm s_{b}\phi (x,\kappa )\frac{\delta \Theta ^{\prime }[\phi (x,\kappa )]}{%
\delta \phi (x,\kappa )},
\end{equation}%
which simplifies to%
\begin{equation}
\frac{d\ln J[\phi ]}{d\kappa }=-\int d^{4}x\sum_{\phi }\pm s_{b}\phi
(x,\kappa )\frac{\delta \Theta ^{\prime }[\phi (x,\kappa )]}{\delta \phi
(x,\kappa )}.  \label{cu}
\end{equation}%
The above expression is nothing else but the expression for an infinitesimal
change in the Jacobian of functional measure. To reach the expression for a
finite Jacobian, it is straightforward to integrate (\ref{cu}) over $\kappa $
within the limits from $0$ to $1$. This leads to the series%
\begin{equation}
\ln J[\phi ]=-\int_{0}^{1}d\kappa \int d^{4}x\sum_{\phi }\pm s_{b}\phi
(x,\kappa )\frac{\delta \Theta ^{\prime }[\phi (x,\kappa )]}{\delta \phi
(x,\kappa )}.
\end{equation}%
Upon making the Taylor expansion of RHS
in $\kappa $ and then integrating over $\kappa $, we find%
\begin{equation}
\ln J[\phi ]=-\left( \int d^{4}x\sum_{\phi }\pm s_{b}\phi (x)\frac{\delta
\Theta ^{\prime }[\phi (x)]}{\delta \phi (x)}\right) .
\end{equation}%

Further simplifications give us a precise expression for the Jacobian of
functional measure under FFBRST transformations: 
\begin{equation}
J[\phi ]={\ \exp \left( -\int d^{4}x\sum_{\phi }\pm s_{b}\phi (x)\frac{%
\delta \Theta ^{\prime }[\phi (x)]}{\delta \phi (x)}\right) }.  \label{J}
\end{equation}%
Here, we notice that, in order to calculate the Jacobian, we have no need of
a local functional $S_{1}[\phi ]$ replacing the Jacobian as $e^{iS_{1}}$ and
satisfying, together with $\Theta ^{\prime }$, certain conditions presented
in Ref. \cite{sdj}. In the FFBRST formulation \cite{sdj}, one first presents
an ansatz for $S_{1}$ in terms of an arbitrary $\kappa $-dependent
parameter; then a physicality condition leads to certain differential
equations with respect to an arbitrary parameter. By satisfying the boundary
conditions, one solves these differential equations to obtain a precise
expression for $S_{1}$.

 Jacobian (\ref{J}) therefore extrapolates the quantum action (within
functional integration) of the theory in (\ref{zen}) as follows: 
\begin{eqnarray}
&&\int \mathcal{D}\phi ^{\prime }e^{iS_{FP}[\phi ^{\prime }]}=\int \mathcal{D}%
\phi \ J[\phi ]e^{iS_{FP}[\phi ]}\nonumber\\
&&=\int \mathcal{D}\phi \ e^{i\left\{
S_{FP}[\phi ]-\int d^{4}x\left( \sum_{\phi }\pm s_{b}\phi \frac{\delta
\Theta ^{\prime }[\phi ]}{\delta \phi }\right) \right\} },
\end{eqnarray}%
which is nothing else but the vacuum functional of the same theory, since
this extra piece does not change the theory on physical grounds, but rather
simplifies various issues in a dramatic way.

\section{Higher-derivative models and FFBRST transformations}

In this section, we present some HD models in the context of FFBRST
description.

\subsection{Higher-derivative Maxwell theory}

The presence of a local gauge symmetry in Maxwell theory requires, as usual,
the introduction of a gauge-fixing term and a compensating Faddeev--Popov
(FP) ghost term to the classical action, resulting in the Faddeev--Popov
quantum action%
\begin{equation}
S_{FP}=\int d^{4}x\left[ -\frac{1}{4}F_{\mu \nu }F^{\mu \nu }-\frac{1}{2}%
\zeta ^{2}(\partial _{\mu }A^{\mu })^{2}-\bar{c}\square c\right] ,
\label{ab}
\end{equation}%
where $\zeta $ is a dimensionless gauge parameter. In the auxiliary field
formulation, the action becomes%
\begin{eqnarray}
S_{FP}&=&\int d^{4}x\left[ -\frac{1}{4}F_{\mu \nu }F^{\mu \nu }+B\partial
_{\mu }A^{\mu }+\frac{1}{2\zeta ^{2}}B^{2}\right.\nonumber\\
&-&\left.\bar{c}\square c\right] ,
\end{eqnarray}%
where $B$ is the Nakanishi--Lautrup field, and $\square =\partial _{\mu
}\partial ^{\mu }$. The Faddeev--Popov action breaks the local gauge
invariance. However, the action $S_{FP}$ remains invariant under a rigid
BRST transformation with a fermionic parameter. The infinitesimal BRST
transformations are%
\begin{eqnarray}
\delta _{b}A_{\mu } &=&-\partial _{\mu }c\Lambda ,\ \ \delta _{b}c=0,  \notag
\\
\delta _{b}\bar{c} &=&B\Lambda ,\ \ \delta _{b}B=0,  \label{brs}
\end{eqnarray}%
where $\Lambda $ is the transformation parameter. There exists a conserved
charge corresponding to the above transformation, which plays an important
role in constructing the physical state space.

An HD version for the quantum action (\ref{ab}) is defined by \cite{bar} 
\begin{eqnarray}
S_{HD} &=&\int d^{4}x\ \left[ -\frac{1}{4}F_{\mu \nu }F^{\mu \nu }-\frac{1}{%
4m^{2}}F_{\mu \nu }\square F^{\mu \nu } \right.  \notag \\
&-&\left. \frac{1}{2}\zeta ^{2}(\partial _{\mu
}A^{\mu })^{2}-\frac{\zeta ^{2}}{2M^{2}}(\partial _{\mu }A^{\mu })\square
(\partial _{\nu }A^{\nu })\right.  \notag \\
&-&\left. \bar{c}\left( 1+\frac{\square }{M^{2}}\right) \square c\right] ,
\label{act}
\end{eqnarray}%
where $m^{2}$ is a dimensional parameter, and $M^{2}$ is a dimensional gauge
parameter. In terms of the auxiliary field $B$, the above expression reads 
\begin{eqnarray}
S_{HD} &=&\int d^{4}x\ \left[ -\frac{1}{4}F_{\mu \nu }F^{\mu \nu }-\frac{1}{%
4m^{2}}F_{\mu \nu }\square F^{\mu \nu }\right.  \notag \\
&+&\left.B\left( 1+\frac{\square }{M^{2}}%
\right) \partial _{\mu }A^{\mu }+\frac{1}{2\zeta ^{2}}B\left( 1+\frac{%
\square }{M^{2}}\right) B\right.  \notag \\
&-&\left. \bar{c}\left( 1+\frac{\square }{M^{2}}\right) \square c\right] .
\label{act1}
\end{eqnarray}%
This HD quantum action is invariant under the same transformations (\ref{brs}%
).

The importance of this HD gauge theory lies in the fact that this model
mimics the model of quantum gravity. For instance, the first term in (\ref%
{act1}) is reminiscent of $\sqrt{-g}R$, and the second term is similar to $%
\sqrt{-g}R^{2}$.

Next, we generalize the BRST transformation according to the above FFBRST
formulation. Following (\ref{a}), (\ref{b}) and (\ref{c}), \ we construct
the FFBRST transformation corresponding to (\ref{brs}), as follows:%
\begin{eqnarray*}
\delta _{b}A_{\mu } &=&-\partial _{\mu }c\Theta \lbrack \phi ],\ \ \delta
_{b}c=0, \\
\delta _{b}\bar{c} &=&B\Theta \lbrack \phi ],\ \ \delta _{b}B=0,
\end{eqnarray*}%
where $\Theta \lbrack \phi ]$ is an arbitrary finite field-dependent
parameter. An explicit choice for the parameter $\Theta \lbrack \phi ]$
produces specific results. To observe the appearance of a higher-derivative
quantum action, we make the following explicit choice: 
\begin{equation}
\Theta ^{\prime }\left[ \phi \right] =\int d^{4}x\left[ \bar{c}\left( \frac{%
\square }{M^{2}}\partial _{\mu }A^{\mu }+\frac{1}{2\zeta ^{2}}\frac{\square 
}{M^{2}}B\right) \right] .
\end{equation}%
Using (\ref{J}), we obtain the Jacobian of functional measure from the above 
$\Theta ^{\prime }$ and find%
\begin{eqnarray}
J[\phi ]&=&\exp \left[ \int d^{4}x\left( B\frac{\square }{M^{2}}\partial _{\mu
}A^{\mu }+\frac{1}{2\zeta ^{2}M^{2}}B\square B\right.\right.  \notag \\
&-&\left.\left. \bar{c}\frac{\square }{M^{2}}%
\square c\right) \right] .
\end{eqnarray}
This Jacobian exhibits BRST-exact HD terms within functional integration. In
other words, HD terms, essential for the quantum action, turn out to be
inherent in the Jacobian for path integral measure under a change of
variables. This justifies a mapping between the Maxwell theory and its HD
version. By computing the Jacobian, one can calculate the HD terms in the
given theory.

\subsection{Higher-derivative theory for non-Abelian vector field}

In this subsection, we extend the above results and use FFBRST
transformations in an HD non-Abelian gauge theory. The action of the theory
is defined by \cite{vo} 
\begin{eqnarray}
S &=&\frac{1}{2}\int d^{d}x\ \mbox{Tr}\left( -F_{\mu \nu }F^{\mu \nu
}+D^{\nu }F_{\nu \mu }D_{\rho }F^{\rho \mu }\right.  \notag \\
&+&\left. \frac{1}{4\xi ^{2}}\partial
^{\mu }\partial A\partial _{\mu }\partial A-\frac{1}{\xi }D^{\nu }F_{\nu \mu
}\partial ^{\mu }\partial A+\bar{\mathcal{F}}^{\mu \nu }\mathcal{F}_{\mu \nu
}\right.  \notag \\
&-&\left. 2i\{\bar{c}^{\mu },c^{\nu }\}F_{\mu \nu }+\frac{1}{\xi }\partial
^{\mu }\bar{c}_{\mu }\partial ^{\nu }c_{\nu }\right) ,  \label{de}
\end{eqnarray}%
where $\xi $ is an arbitrary gauge parameter. Here, the Yang--Mills
covariant derivative is defined by $D_{\mu }=\partial _{\mu }+g[A_{\mu
},\bullet ]$; $\mathcal{F}_{\mu \nu }$ and $\bar{\mathcal{F}}_{\mu \nu }$
are the field strengths for the fields $c_{\mu }$ and $\bar{c}_{\mu }$,
respectively. The above action is invariant under the following rigid
fermionic symmetry: 
\begin{eqnarray}
&&\delta _{b}A_{\mu }^{a}=-c_{\mu }^{a}\Lambda ,\ \ \delta _{b}c_{\mu
}^{a}=0,\nonumber\\
&& \delta _{b}\bar{c}_{\mu }^{a}=\left( D^{\nu ab}F_{\nu \mu }^{b}+%
\frac{1}{2\xi }\partial _{\mu }\partial ^{\nu }A_{\nu }^{a}\right) \Lambda .
\end{eqnarray}
Using the auxiliary field $b_{\mu }^{a}$, we present the action (\ref{de})
in the form%
\begin{eqnarray*}
S &=&\frac{1}{2}\int d^{d}x\ \mathrm{Tr}\left[ -F_{\mu \nu }F^{\mu \nu
}+b^{\mu }\left( D^{\nu }F_{\nu \mu }\right. \right.\\
&+&\left.\left. \frac{1}{2\xi }\partial _{\mu
}\partial ^{\nu }A_{\nu }-\frac{1}{2}b_{\mu }\right) +\bar{\mathcal{F}}^{\mu
\nu }\mathcal{F}_{\mu \nu }\right. \\
&-&\left. 2i\{\bar{c}^{\mu },c^{\nu }\}F_{\mu \nu }+\frac{1}{\xi }\partial
^{\mu }\bar{c}_{\mu }\partial ^{\nu }c_{\nu }\right] ,
\end{eqnarray*}%
which is invariant under the following off-shell nilpotent BRST
transformations: 
\begin{equation}
\delta _{b}A_{\mu }^{a}=-c_{\mu }^{a}\Lambda ,\ \ \delta _{b}c_{\mu
}^{a}=0,\ \ \delta _{b}\bar{c}_{\mu }^{a}=b_{\mu }^{a}\Lambda ,\ \ \delta
_{b}b_{\mu }^{a}=0.
\end{equation}%
This structure has been discussed in topological quantum field theories \cite%
{brook}. These transformations are generalized by making the transformation
parameter finite and field-dependent:%
\begin{eqnarray*}
&&\delta _{b}A_{\mu }^{a}=-c_{\mu }^{a}\Theta \lbrack \phi ],\ \ \delta
_{b}c_{\mu }^{a}=0,\ \ \delta _{b}\bar{c}_{\mu }^{a}=b_{\mu }^{a}\Theta
\lbrack \phi ],\nonumber\\
&& \delta _{b}b_{\mu }^{a}=0,
\end{eqnarray*}%
where the finite parameter is constructed explicitly from the infinitesimal
field-dependent parameter 
\begin{eqnarray}
\Theta ^{\prime }[\phi ]&=&\frac{1}{2}\int d^{d}x\ \mathrm{Tr}\left[ \bar{c}%
^{\mu }\left( D^{\nu }F_{\nu \mu }+\frac{1}{2\xi }\partial _{\mu }\partial
^{\nu }A_{\nu }\right. \right.\nonumber\\
&+&\left.\left.\frac{1}{2}b_{\mu }\right) \right] .  \label{hu}
\end{eqnarray}%
The Jacobian of functional measure under the FFBRST transformation with a
parameter constructed by (\ref{hu}) reads as follows:%
\begin{eqnarray}
J[\phi ] &=&\exp \left\{ \frac{1}{2}\int d^{d}x\ \mathrm{Tr}\left[ b^{\mu
}\left( D^{\nu }F_{\nu \mu }+\frac{1}{2\xi }\partial _{\mu }\partial ^{\nu
}A_{\nu }\right. \right.\right.\nonumber\\
&-&\left.\left.\left.\frac{1}{2}b_{\mu }\right) +\bar{\mathcal{F}}^{\mu \nu }\mathcal{F}%
_{\mu \nu }-2i\{\bar{c}^{\mu },c^{\nu }\}F_{\mu \nu }\right. \right.  \notag \\
&+&\left. \left. \frac{1}{\xi }%
\partial ^{\mu }\bar{c}_{\mu }\partial ^{\nu }c_{\nu }\right] \right\} .
\label{gy}
\end{eqnarray}%
Now, we can see that under FFBRST transformations with a specific parameter
one can produce an HD action for the non-Abelian theory in question. This
also justifies the validity of our approach in non-Abelian gauge theories.
Consequently, using FFBRST transformations, one can generate appropriate HD
terms which allow one to get rid of UV divergencies. Since the HD theory is
BRST-invariant, the unitarity problem associated with HD theories can be
overcome.

\subsection{Higher-derivative gravity}

In this subsection, we examine FFBRST transformations in HD gravity. To this
end, we start with a general fourth-order gravity action in curved spacetime 
\cite{anto}, 
\begin{eqnarray}
S_{g}&=&\int d^{4}x\sqrt{-g}\left[ -\frac{1}{\alpha ^{2}}\left( R_{\mu \nu
}R^{\mu \nu }-\frac{1}{3}R^{2}\right) +\beta R^{2}\right.\nonumber\\
&+&\left.\frac{\gamma }{\zeta ^{2}}%
R\right] .  \label{hd}
\end{eqnarray}
In the weak limit, we decompose the metric into a fixed metric $\eta ^{\mu
\nu }$ and fluctuations $h^{\mu \nu }$, as follows:%
\begin{equation}
\sqrt{-g}g^{\mu \nu }=\eta ^{\mu \nu }+\alpha \zeta h^{\mu \nu }.
\end{equation}%
The action (\ref{hd}) is invariant under the following general gauge transformation: 
\begin{equation}
\delta h^{\mu \nu }=D_{\rho }^{\mu \nu }\omega ^{\rho },
\end{equation}%
where the manifest expression for the covariant derivative of the vector
parameter $\omega ^{\rho }$ is given by%
\begin{eqnarray}
D_{\rho }^{\mu \nu }\omega ^{\rho }&=&\partial ^{\mu }\omega ^{\nu }+\partial
^{\nu }\omega ^{\mu }-\eta ^{\mu \nu }\partial _{\rho }\omega ^{\rho
}+\alpha \zeta (\partial _{\rho }\omega ^{\mu }h^{\rho \nu }\nonumber\\
&+&h^{\rho \mu
}\partial _{\rho }\omega ^{\nu }-\partial _{\rho }h^{\mu \nu }\omega ^{\rho
}-h^{\mu \nu }\partial _{\rho }\omega ^{\rho }).
\end{eqnarray}
According to conventional quantization, one introduces gauge-fixing in order
to remove the redundant degrees of freedom. Here, we choose the familiar
harmonic (De Donder) gauge%
\begin{equation}
\partial _{\nu }h^{\mu \nu }=0.
\end{equation}%
Then the gauge-fixing term in the action is quadratic in derivatives:%
\begin{equation}
S_{gf}=-\frac{1}{2}\int d^{4}x\ (\partial _{\nu }h^{\mu \nu })^{2}.
\end{equation}%
This implies that it is not every part of the graviton propagator that
behaves as $($momentum$)^{-4}$ for large momenta, leading thereby to some UV
divergences. This complication is easily overcome by introducing
gauge-fixing terms with four or more derivatives \cite{anto}: 
\begin{equation}
S_{gf}=-\frac{1}{2}\int d^{4}x\ [\hat{e}(\square )\partial _{\nu }h^{\mu \nu
}]^{2},
\end{equation}%
where $\hat{e}(\square )=b_{1}\square +b_{2}$, with $b_{1}$ and $b_{2}$
being constant. Using the Nakanishi--Lautrup field $B_{\mu }$, one presents
the linearized gauge-fixing term as%
\begin{equation}
S_{gf}=\int d^{4}x\left[ \frac{1}{2}(B_{\mu })^{2}-B_{\mu }b_{1}\square
\partial _{\nu }h^{\mu \nu }-B_{\mu }b_{2}\partial _{\nu }h^{\mu \nu }\right].
\end{equation}%
The compensating ghost term within functional integration is given by 
\begin{eqnarray*}
S_{gh} &=&\int d^{4}x\left\{ \bar{c}_{\mu }b_{1}\square \partial _{\nu }%
\left[ \partial ^{\mu }c^{\nu }+\partial ^{\nu }c^{\mu }-\eta ^{\mu \nu
}\partial _{\rho }c^{\rho }\right.\right.\nonumber\\
&+&\left.\left.\alpha \zeta \left( \partial _{\rho }c^{\mu
}h^{\rho \nu }+h^{\rho \mu }\partial _{\rho }c^{\nu }-\partial _{\rho
}h^{\mu \nu }c^{\rho }-h^{\mu \nu }\partial _{\rho }c^{\rho }\right) \right]
\right. \\
&+&\left. \bar{c}_{\mu }b_{2}\partial _{\nu }\left[ \partial ^{\mu }c^{\nu
}+\partial ^{\nu }c^{\mu }-\eta ^{\mu \nu }\partial _{\rho }c^{\rho }+\alpha
\zeta \left( \partial _{\rho }c^{\mu }h^{\rho \nu }
\right.\right.\right.\nonumber\\
&+&\left.\left. \left. h^{\rho \mu }\partial
_{\rho }c^{\nu }-\partial _{\rho }h^{\mu \nu }c^{\rho }-h^{\mu \nu }\partial
_{\rho }c^{\rho }\right) \right] \right\} .
\end{eqnarray*}%
The FP quantum action $S_{g}+S_{gf}+S_{gh}$ admits BRST invariance under%
\begin{eqnarray}
\delta _{b}h^{\mu \nu } &=&D_{\rho }^{\mu \nu }c^{\rho }\Lambda ,\ \ \delta
_{b}c^{\mu }=-\zeta \partial _{\nu }c^{\mu }c^{\nu }\Lambda ,  \notag \\
\delta _{b}\bar{c}^{\mu } &=&-B^{\mu }\Lambda \ \ \delta _{b}B^{\mu }=0.
\label{xy}
\end{eqnarray}%
Using these symmetry transformations, one can compute a conserved (BRST)
charge which annihilates the physical states in the total state space and
helps one to establish unitarity in the theory.

Following Section II, we now construct the FFBRST transformations corresponding
to (\ref{xy}), namely,%
\begin{eqnarray}
\delta _{b}h^{\mu \nu } &=&D_{\rho }^{\mu \nu }c^{\rho }\ \Theta \lbrack
\phi ],\ \ \delta _{b}c^{\mu }=-\zeta \partial _{\nu }c^{\mu }c^{\nu
}\Lambda \ \Theta \lbrack \phi ],  \notag \\
\delta _{b}\bar{c}^{\mu } &=&-B^{\mu }\Lambda \ \Theta \lbrack \phi ]\ \
\delta _{b}B^{\mu }=0.
\end{eqnarray}%
where $\Theta \lbrack \phi ]$ is an arbitrary finite field-dependent
parameter. This parameter admits any value; however, in the present case we
assign it to the value which is derived from%
\begin{equation}
\Theta ^{\prime }\left[ \phi \right] =\int d^{4}x\ \bar{c}_{\mu
}(b_{1}\square \partial _{\nu }h^{\mu \nu }).  \label{the}
\end{equation}%
Using (\ref{J}) and (\ref{the}), we calculate the Jacobian of functional
measure:%
\begin{eqnarray*}
J[\phi ] &=&\exp \left\{ \int d^{4}x\left[ -B_{\mu }b_{1}\square \partial
_{\nu }h^{\mu \nu }+\bar{c}_{\mu }b_{1}\square \partial _{\nu }\left[
\partial ^{\mu }c^{\nu }\right. \right. \right. \\
&+&\left. \left.\left.  \partial ^{\nu }c^{\mu }-\eta ^{\mu \nu }\partial
_{\rho }c^{\rho }+ \alpha \zeta  ( \partial _{\rho }c^{\mu }h^{\rho \nu
}+h^{\rho \mu }\partial _{\rho }c^{\nu }\right. \right. \right. \\
&-&\left. \left.\left. \partial _{\rho }h^{\mu \nu
}c^{\rho }-h^{\mu \nu }\partial _{\rho }c^{\rho } ) \right] \right] \right\} .
\end{eqnarray*}%
So, we can see that this parameter renders FFBRST transformations a source
of HD terms in the quantum action of gravity. This proves our treatment to
be valid also in the case of gravity. It is well known that the action (\ref%
{hd}) is renormalizable by power counting, and, in fact, this
renormalizability has been demonstrated in Ref. \cite{ste}. More
importantly, this theory is asymptotically free \cite{tom,fra}. The
renormalizability and asymptotic freedom are entirely due to the HD terms.
However, there is still redundancy in physical degrees of freedom. To remove
it, one needs a higher-derivative quantum action, which can be generated
using the FFBRST mechanism with suitable HD terms in the theory through the
Jacobian.

\section{Conclusion}

HD field theories are of interest, since they play an important role in
understanding the fundamental interactions of Nature. Incidentally, the
theory of gravity, as we know it today, is an effective theory, and the
usual Einstein--Hilbert action should be supplemented with corrections
involving higher powers in the curvature tensor. This is supported by string
theory or by conformal anomalies present in all quantum field theories
coupled to gravity. From the practical viewpoint, HD gravity endows the
effective potential and phase transitions of scalar fields with a wealth of
astrophysical and cosmological properties.

In this paper, we have generalized rigid BRST transformations by allowing
the transformation parameter to be finite and field-dependent. The
expression for the Jacobian presented here has a more solid derivation
basis. To calculate the Jacobian, we do not need any local functional
satisfying some initial conditions and differential equations. Here, the
Jacobian depends on an arbitrary infinitesimal field-dependent parameter.
For a given value of the field-dependent parameter, one can easily compute
the Jacobian of functional measure under FFBRST transformations. We have
implemented such FFBRST transformations in different HD models. For
instance, we employed the FFBRST formalism first in Maxwell theory and found
that, for a particular value of the field-dependent parameter, the Jacobian
is the source of HD terms in the BRST-exact part of the theory. At the same
time, BRST symmetry, in its finite field-dependent form, makes it possible
to provide independence for the S-matrix from any particular HD
gauge-fixing. That is to say, such a BRST transformation actually preserves
the S-matrix and transforms a quantum theory into an equivalent one. To
extend this result, we have further studied FFBRST transformations in a
non-Abelian theory and in quantum gravity. Here, remarkably, we have found
the previous general results to hold true for these theories as well. Thus,
we have mapped different HD theories to the BRST-exact parts of these
theories. The HD terms in the quantum action have been generated in a
precise form through the Jacobian of functional measure. So, we conclude
that the Jacobian of functional measure plays a key role in this treatment.

Even though an HD action is renormalizable by power counting, and, in fact,
this renormalizability has been established in full generality, the nature
of HD terms in the quantum action requires that one remove some redundancies
in gauge degrees of freedom, which are generated through the BRST
transformations. The present study may be of help in dealing with a theory
having UV-divergent terms. It will be interesting to use the results of this
paper to establish renormalizability in some models by getting rid of UV
divergences.

Recently, a concept of the Very Special Relativity (VSR) has been suggested 
\cite{ko}. It is based on the idea that the laws of physics need not be
invariant under the full Lorentz group, but rather under its subgroups,
which still preserves the basic SR elements, such as the constancy of the
speed of light. VSR has been under active investigation by many researchers 
\cite{9,10,mu,12,16,sud1}. It will be interesting to study FFBRST and HD
theories in the VSR context.

\textbf{Acknowledgement} The work of P.Yu.M. was supported by the Tomsk
State University Competitiveness Improvement Program. SU acknowledges the financial 
support from Indian Institute of Technology Kharagpur, India under PDF program.


\begin{thebibliography}{99}
\bibitem{stelle} K.~S. ~Stelle, Phys. Rev. D {16}, 953 (1977).

\bibitem{gib} D. G. Boulware and S. Deser. Phys. Rev. D 6, 3368 (1972); G.
Gibbons, Phys. Rev. Lett. 64, 123 (1990).

\bibitem{anto} I. Antoniadis and E. T. Tomboulis, Phys. Rev. D 33, 2756
(1986).

\bibitem{pisarski} R. D. Pisarski, Phys. Rev. D {\ {34}}, 670 (1986).

\bibitem{nesterenko} V. V. Nesterenko, J. Phys. A {\ {\ 22}}, 1673 (1989).

\bibitem{plyuschay1} M. S. Plyushchay, Mod. Phys. Lett. A {3}, 1299 (1988);

\bibitem{plyuschayA} M. S. Plyushchay, Int. J. Mod. Phys. A {4}, 3851 (1989).

\bibitem{plyuschay2} M. S. Plyushchay, Mod. Phys. Lett. A {4}, 837 (1989);
Phys. Lett. B {\ 243}, 383 (1990).

\bibitem{ramos} E. Ramos, J. Roca, Nucl. Phys. B {\ {\ 436}}, 529 (1995).

\bibitem{ramos2} E. Ramos, J. Roca, Nucl. Phys. B {452 }, 705 (1995).

\bibitem{BMP} R. ~Banerjee, P. ~Mukherjee, B. ~Paul, JHEP {\ 1108}, 085
(2011).

\bibitem{rbs} R. Banerjee, B. Paul and S. Upadhyay, Phys. Rev. D 88, 065019
(2013).

\bibitem{podolsky1} B.~Podolsky, Phys. Rev. {\ 62}, 68 (1942).

\bibitem{podolsky2} B.~Podolsky, C.~Kikuchi, Phys. Rev. {\ 65}, 228 (1944).

\bibitem{Iliopoulos} J. Iliopoulos, B. Zumino, Nucl. Phys. B {76}, 310
(1974).

\bibitem{Gama} F. S. Gama, M. Gomes, J. R. Nascimento, A. Yu. Petrov, A. J.
da Silva. Phys. Rev. D {84}, 045001 (2011).

\bibitem{clz} C.~S.~Chu, J.~Lukierski, W.~J.~Zakrzewski, Nucl. Phys. B {\ 632%
}, 219 (2002).

\bibitem{plyuschay6} P. D. Alvarez, J. Gomis, K. Kamimura, M. S. Plyushchay,
Phys. Lett. B {659}, 906 (2008).

\bibitem{neupane} I. P. Neupane, JHEP {\ 09}, 040 (2000).

\bibitem{nojiri4} S. Nojiri, S. D. Odintsov, S. Ogushi, Phys. Rev. D {65},
023521 (2001).

\bibitem{reyes} C. ~M. ~Reyes, Phys. Rev. D {80}, 105008 (2009).

\bibitem{MP} P. ~Mukherjee, B. ~Paul, Phys. Rev. D {85}, 045028 (2012).

\bibitem{plyuschay3} M. S. Plyushchay, Nucl. Phys. B {362 }, 54 (1991).

\bibitem{plyuschay4} Peter A. Horv\' athy, M. S. Plyushchay, JHEP {0206},
033 (2002).

\bibitem{plyuschay5} M. S. Plyushchay, Electron. J. Theor. Phys. {3N10}, 17
(2006).

\bibitem{plyuschay7} M. S. Plyushchay, Phys. Lett. B {262}, 71 (1991).

\bibitem{cordero} R. Cordero, A. Molgado, E. Rojas, Class. Quantum Grav. {28}%
, 065010 (2011).

\bibitem{paul} B. Paul, Phys. Rev. D {\ 87}, 045003 (2013). 

\bibitem{accioly} A. J. Accioly, Revlsta Brasllelra de Flslca {18 }, 593
(1988).

\bibitem{soti} T. P. Sotiriou, V. Faraoni, Rev. Mod. Phys. {82}, 451 (2010).

\bibitem{gullu} I. Gullu, T. C. Sisman, B. Tekin, Phys. Rev. D {81}, 104017
(2010).

\bibitem{ohta} N. Ohta, Class. Quantum Grav. {29}, 015002 (2012).

\bibitem{polya} A. M. Polyakov, Nucl. Phys. B {268}, 406 (1986).

\bibitem{elie} D. A. Eliezer, R. P. Woodard, Nucl. Phys. B {325}, 389 (1989).

\bibitem{brst} C. Becchi, A. Rouet, R. Stora, Annals Phys. \textbf{98}
(1974) 287.

\bibitem{tyu} I. V. Tyutin, LEBEDEV-\textbf{75-39} (1975).

\bibitem{sudu} S. Upadhyay, EPL 103, 61002 (2013); Phys. Lett. B 723, 470
(2013); Eur. Phys. J. C 74, 2737 (2014).

\bibitem{wei} S. Weinberg, \textit{\ The quantum theory of fields, Vol-II:
Modern applications}, Cambridge, UK Univ. Press (1996).

\bibitem{sdj} S. D. Joglekar and B. P. Mandal, Phys. Rev. D. 51, 1919 (1995).

\bibitem{sdj1} S. D. Joglekar and B. P. Mandal, Int. J. Mod. Phys. A 17,
1279 (2002).

\bibitem{susk} S. Upadhyay, S. K. Rai and B. P. Mandal, J. Math. Phys. {52}, 
{022301} (2011).

\bibitem{gri} V. N. Gribov, Nucl. Phys. B 139, 1 (1978).

\bibitem{zwan} D. Zwanziger, Nucl. Phys. B 323, 513 (1989).

\bibitem{zwan1} D. Zwanziger, Nucl. Phys. B 399, 477 (1993).

\bibitem{sb} S. Upadhyay and B. P. Mandal, Eur. Phys. J. C 75, 327 (2015);
Int. J. Theor. Phys. 55, 1 (2016); Phys. Lett. B 744, 231 (2015); Prog.
Theor. Exp. Phys. 053B04 (2014); Eur. Phys. J. {C 72}, 2065 (2012); Annls.
Phys. {\ 327}, 2885 (2012); Eur. Phys. Lett. {\ 93}, 31001 (2011); Mod.
Phys. Lett. {A 25}, {\ 3347} (2010).

\bibitem{smm} S. Upadhyay, M. K. Dwivedi and B. P. Mandal, Int. J. Mod.
Phys. A 30, 1550178 (2015); Int. J. Mod. Phys. A 28, 1350033 (2013).

\bibitem{bl} J.F Bagger and N. Lambert, Phys. Rev. D 75, 045020 (2007);
Phys. Rev. D 77, 065008 (2008); JHEP 0802, 105 (2008).

\bibitem{g} A. Gustavsson, Nucl. Phys. B 811, 66 (2009).

\bibitem{abjm} O. Aharony, O. Bergman, D. L. Jafferis and J. Maldacena,
JHEP. 0810, 091 (2008).

\bibitem{fs} M. Faizal, B. P. Mandal and S. Upadhyay, Phys. Lett. B 721, 159
(2013).

\bibitem{sudd} S. Upadhyay and D. Das, Phys. Lett. B 733, 63 (2014).

\bibitem{fsm} M. Faizal, S. Upadhyay and B. P. Mandal, Phys. Lett. B 738,
201 (2014); Int. J. Mod. Phys. A 30, 1550032 (2015); S. Upadhyay, M. Faizal
and P. A. Ganai, Int. J. Mod. Phys. A 30, 1550185 (2015).

\bibitem{bisu} S. Upadhyay and B. Paul, Eur. Phys. J. C  76, 394 (2016); S.
Upadhyay, M. Oksanen and R. Bufalo, arXiv:1510.00188 [hep-th]; M. Faizal, S.
Upadhyay and B. P. Mandal, Eur. Phys. J. C 76, 189 (2016).

\bibitem{ssb} B. P. Mandal, S. K. Rai and S. Upadhyay, EPL {\ 92}, {21001}
(2010).

\bibitem{sud001} S. Upadhyay, Phys. Lett. B 740, 341 (2015); Annls. Phys.
356, 299 (2015); Mod. Phys. Lett. A 30,1550072 (2015); Annls. Phys. 340, 110
(2014); Annls. Phys. 344, 290 (2014); EPL 105, 21001 (2014); EPL 104, 61001
(2013); Phys. Lett. B 727, 293 (2013).

\bibitem{ht} M. Henneaux, C. Teitelboim, \textit{\ Quantization of gauge
systems}, Princeton, USA: Univ. Press (1992).

\bibitem{rs} R. Banerjee and S. Upadhyay, Phys. Lett. B 734, 369 (2014).

\bibitem{ale} P.~Y.~Moshin and A.~A.~Reshetnyak, {\ Nucl. Phys. B 888}, 92
(2014).

\bibitem{ale1} P.~Y.~Moshin and A.~A.~Reshetnyak, {\ Int. J. Mod. Phys. A {30%
}, 1550021 (2015)}.

\bibitem{mos1} P.~Y.~Moshin and A.~A.~Reshetnyak, {Phys. Lett. B {\ 739},
110 (2014)}.
\bibitem{asm} S. Upadhyay, A. Reshetnyak and B. P. Mandal, Eur. Phys. J. C  76, 391 (2016).

\bibitem{mos} P.~Y.~Moshin and A.~A.~Reshetnyak, Int. J. Mod. Phys. A 31, 1650111 (2016).

\bibitem{bar} A. Bartoli and J. Julve, Nucl. Phys. B 425, 277 (1994).

\bibitem{vo} V. O. Rivelles, Phys. Lett. B 577, 137 (2003).

\bibitem{brook} R. Brooks, D. Montano and J. Sonnenschein, Phys. Lett. B
214, 91 (1988).

\bibitem{ste} K. S. Stelle, Phys. Rev. D 16, 953 (1977).

\bibitem{tom} E. T. Tomboulis, Phys. Lett. B 97, 77 (1980).

\bibitem{fra} F. S. Fradkin and A. A. Tseytlin, Nucl. Phys. B 201, 469
(1982).

\bibitem{ko} A. G. Cohen and S. L. Glashow, Phys. Rev. Lett. 97, 021601
(2006).

\bibitem{9} A. G. Cohen and D. Z. Freedman, J. High Energy Phys. 07 039,
(2007); J. Vohanka, Phys. Rev. D 85, 105009 (2012).

\bibitem{10} G.W. Gibbons, J. Gomis, and C. N. Pope, Phys. Rev. D 76, 081701
(2007).

\bibitem{mu} W. Muck, Phys. Lett. B 670, 95 (2008).

\bibitem{12} E. Alvarez and R. Vidal, Phys. Rev. D 77, 127702 (2008).

\bibitem{16} J. Alfaro and V. O. Rivelles, Phys. Rev. D 88, 085023 (2013).

\bibitem{sud1} S. Upadhyay, Eur. Phys. J. C 75, 593 (2015); S. Upadhyay and P. K. Panigrahi, Nucl. Phys. B 915, 168 (2017).
\end{thebibliography}
\end{document}